\documentclass{Interspeech}
\usepackage{svg,multirow,adjustbox}

\interspeechcameraready

\title{Pseudo Labels-based Neural Speech Enhancement for the AVSR Task in the MISP-Meeting Challenge}

\author[affiliation={1}]{Longjie}{Luo}
\author[affiliation={2}]{Shenghui}{Lu}
\author[affiliation={1*}]{Lin}{Li}
\author[affiliation={2*}]{Qingyang}{Hong}

\affiliation{School of Electronic Science and Engineering}{Xiamen University}{China}
\affiliation{School of Informatics}{Xiamen University}{China}
\email{\{lilin,qyhong\}@xmu.edu.cn}

\usepackage{comment}

\begin{document}

\maketitle

\makeatletter
\def\@makefnmark{} 
\footnotetext{* Corresponding author.}
\def\@makefnmark{\hbox{\@textsuperscript{\normalfont\@thefnmark}}} 
\makeatother

\begin{abstract}
This paper presents our system for the MISP-Meeting Challenge Track 2. The primary difficulty lies in the dataset, which contains strong background noise, reverberation, overlapping speech, and diverse meeting topics. To address these issues, we (a) designed G-SpatialNet, a speech enhancement (SE) model to improve Guided Source Separation (GSS) signals; (b) proposed TLS, a framework comprising \textbf{t}ime alignment, \textbf{l}evel alignment, and \textbf{s}ignal-to-noise ratio filtering, to generate signal-level pseudo labels for real-recorded far-field audio data, thereby facilitating SE models' training; and (c) explored fine-tuning strategies, data augmentation, and multimodal information to enhance the performance of pre-trained Automatic Speech Recognition (ASR) models in meeting scenarios. Finally, our system achieved character error rates (CERs) of 5.44\% and 9.52\% on the Dev and Eval sets, respectively, with relative improvements of 64.8\% and 52.6\% over the baseline, securing second place.
\end{abstract}

\section{Introduction}
With the advancement of multimodal technology, Audio-Visual Speech Recognition (AVSR) has attracted increasing attention. Previous Multimodal Information based Speech Processing (MISP) challenges released a large-scale Mandarin audio-visual conversational dataset for home-TV scenarios \cite{chen2022first,wang2023multimodal,wu2024multimodal}, promoting progress in far-field AVSR.
The MISP-Meeting Challenge focuses on meeting scenarios and releases the MISP-Meeting dataset, a large-scale Mandarin audio-visual meeting corpus.
The dataset contains 125 hours of audio-visual data, including far-field audio captured by an 8-channel microphone array, high-quality close-talk recordings, and video footages of the speakers.
The organizers increases the challenge difficulty by diversifying meeting room acoustics, expanding the range of meeting topics, and raising the proportion of overlapping speech. 
And the organizers allow the use of publicly available pre-trained models, offering valuable guidance for advancing pre-trained Automatic Speech Recognition (ASR) technology in Chinese meeting scenarios.
We focus on MISP-Meeting Challenge Track 2, an AVSR task, aiming to explore: data-driven front-end algorithms in real-world complex acoustic environments, fine-tuning strategies for pre-trained ASR models in Chinese meeting scenarios, effectiveness of data augmentation and multimodal information for this task.

\begin{figure}
    \centerline{\includegraphics[width=\columnwidth]{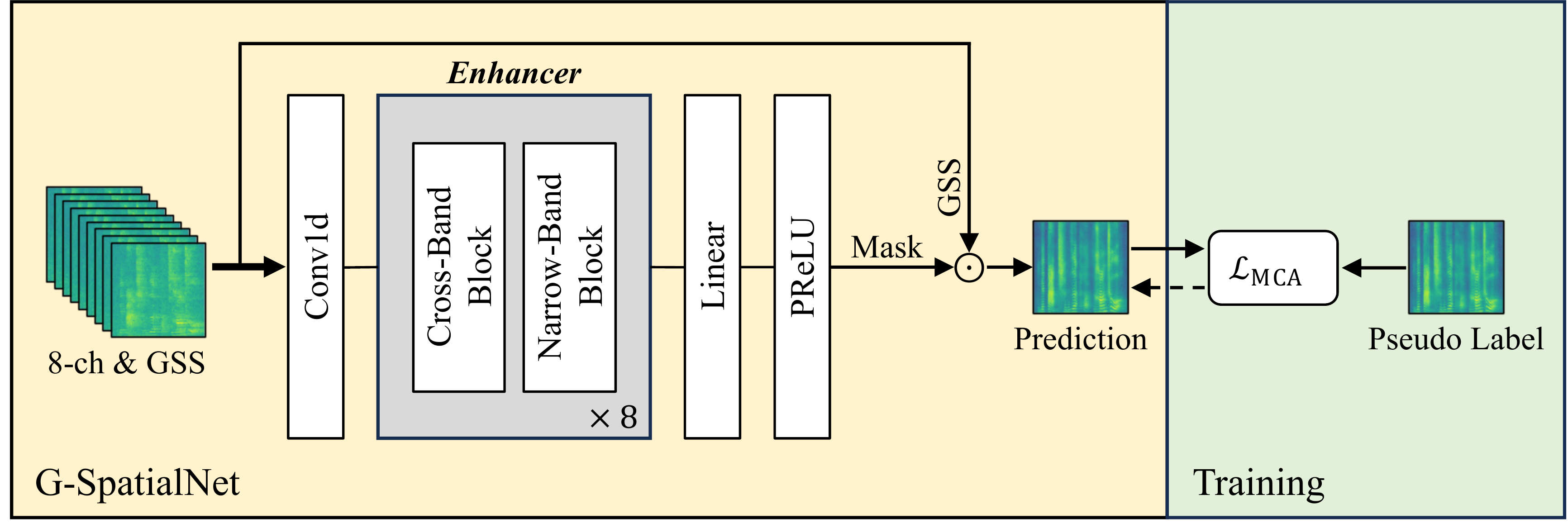}}
    \caption{Proposed G-SpatialNet model and pseudo labels-based training on real-recorded data.}
    \label{figure4}
    \vspace{-10pt} 
\end{figure}

For the front-end, Guided Source Separation (GSS) \cite{boeddeker2018front, raj23_interspeech} has demonstrated outstanding performance in far-field multi-speaker conversational scenarios. In challenges like CHiME-6/-7 \cite{watanabe2020chime,cornell2023chime}, MISP2021/2022 \cite{chen2022first,wang2023multimodal}, nearly all teams adopted GSS as their only front-end algorithm. However, as a pure signal processing approach, GSS cannot leverage large-scale training data, and its enhanced speech still contains noticeable noise and reverberation. To address these issues, we introduce an additional speech enhancement (SE) model called G-SpatialNet. On the other hand, since reference clean speech is unavailable in real-world scenarios, previous studies \cite{wang2023multimodal,cornell2024chime,vinnikov2024notsofar} mainly rely on simulated data to train SE models. However, this brings domain mismatch problem between simulated and real-recorded data, which significantly degrades models performance in real-world scenarios \cite{leglaive2023chime,10566014,wang2024unssor}. To mitigate this issue, as shown in Fig. \ref{figure4} and Fig. \ref{figure1}, we propose a novel pseudo labels-based training method. Specifically, we propose TLS, a framework consisting of \textbf{t}ime alignment, \textbf{l}evel alignment, and \textbf{s}ignal-to-noise ratio filtering, to generate signal-level pseudo labels for real-recorded far-field audio data for the first time. We then train the models directly on real-recorded data by optimizing the proposed magnitude constraint adjustable (MCA) loss.

For the back-end, we investigate the impact of the visual modality. In the Chinese meeting scenario of the dataset, we employ Paraformer \cite{paraformer}, an industrial-grade Mandarin pre-trained ASR model, as our back-end. Given the limited research on applying pre-trained models to Chinese meeting scenarios, a key challenge in improving performance is preserving the pre-trained model's capabilities while adapting it to the MISP-Meeting data distribution. To address this, we explore the effectiveness of data augmentation and identify the optimal fine-tuning strategy.

\section{Front-end}
In our front-end, we first employ GSS with the oracle diarization to extract each speaker's speech signal, followed by a dedicated SE model called G-SpatialNet (Section \ref{section:Speech enhancement model}) for further signal enhancement. To enable direct training on real-world meeting data, we propose a novel pseudo labels-based approach, which includes the TLS framework (Section \ref{section:Pseudo labels estimation}) for pseudo labels generation and MCA loss (\ref{section:Loss function}) for model training. 

\subsection{Guided source separation}
\label{section:Guided source separation}
We followed the GSS implementation in \cite{raj23_interspeech}, while using the same parameters as the official baselines\footnote{\url{https://github.com/mispchallenge/MISP2025-AVSR-baseline/}}. The GSS algorithm consists of multi-channel weighted prediction error (WPE) \cite{nakatani2008blind,nakatani2010speech} for dereverberation, complex Aagular Central Gaussian Mixture
Models (cACGMMs) \cite{ito2016complex} guided by the diarization information for mask estimation, and mask-based minimum-variance distortionless
response (MVDR) beamformer \cite{souden2009optimal,erdogan2016improved} for spatial filtering, optionally followed by time-frequency (T-F) masking.

\subsection{Pseudo labels-based neural speech enhancement}

\subsubsection{G-SpatialNet}
\label{section:Speech enhancement model}
SpatialNet \cite{quan2024spatialnet}, a recently proposed model for multi-channel joint speech separation, denoising, and dereverberation, demonstrates state-of-the-art (SOTA) or near-SOTA performance across various SE sub-tasks. 

To enhance GSS signals, we design G-SpatialNet, an adapted version of SpatialNet. As shown in Fig. \ref{figure4}, the main modifications in G-SpatialNet include: (a) replacing input feature from complex to magnitude spectrograms, as the ASR back-ends just utilize magnitude information; (b) following \cite{luo2024xmuspeech}, we stack the magnitude spectrograms of both GSS and the raw 8-channel microphone array signal along the channel dimension as the model input; and (c) adopting ideal amplitude mask (IAM) \cite{erdogan2015phase} as the training target instead of direct mapping to mitigate the speech distortion.

\begin{figure}
    \centerline{\includegraphics[width=\columnwidth]{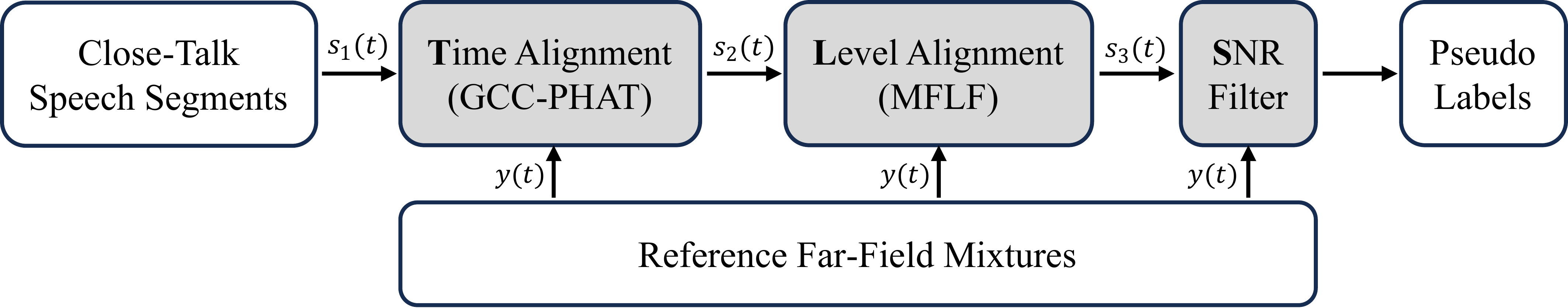}}
    \caption{Proposed TLS framework for pseudo labels estimation. In this paper, the “Reference Far-Field Mixtures” are designated as GSS signals.}
    \label{figure1}
    \vspace{-15pt} 
\end{figure}

\subsubsection{Pseudo labels estimation}
\label{section:Pseudo labels estimation}
In far-field multi-speaker conversational scenarios, such as CHiME \cite{watanabe2020chime,cornell2023chime,barker2018fifth}, M2MeT \cite{yu2022m2met}, and MISP \cite{chen2022first,wang2023multimodal,wu2024multimodal}, the close-talk speech signal of each speaker is recorded in addition to the far-field microphone array signals for manual transcription. Since close-talk recordings typically have high-quality, it is reasonable to
use them for pseudo labels estimation. However, the close-talk recordings are time- and level-misaligned with the target clean speech (direct sound in this paper) due to signal delay and attenuation from long-distance propagation. To address the issue, we propose a novel framework called TLS. As shown in Fig. \ref{figure1}, the TLS framework comprises \textbf{t}ime alignment, \textbf{l}evel alignment and \textbf{s}ignal-to-noise ratio (SNR) filtering.

\begin{table}[]
    \caption{Comparison of ASR performance and DNSMOS P.835 scores for close-talk speech segments (CTSS), GSS signals, and pseudo labels (PL) on the MISP-Meeting development set. Note that for fairness, Paraformer\protect\footnotemark[2] as well as Whisper\protect\footnotemark[3] use official pre-trained parameters.}
    \begin{adjustbox}{width=\columnwidth}
    \renewcommand{\arraystretch}{1.2}
    \begin{tabular}{cccccc}
    \toprule
    \multirow{2}{*}{\textbf{Data}} & \multicolumn{2}{c}{\textbf{CER(\%) $\downarrow$}} & \multicolumn{3}{c}{\textbf{DNSMOS P.835} \cite{reddy2022dnsmos}} \\ \cmidrule(lr){2-3} \cmidrule(l){4-6}
    & \textbf{Paraformer} \cite{paraformer} & \textbf{Whisper} \cite{radford2023robust} & \textbf{SIG $\uparrow$} & \textbf{BAK $\uparrow$} & \textbf{OVRL $\uparrow$} \\ 
    \toprule
    CTSS & 4.26 & 5.32 & 3.35 & 3.49 & 2.79 \\
    GSS & 7.03 & 9.49 & 2.48 & 2.26 & 1.89 \\ 
    \hline
    PL & 4.71 & 5.94 & 3.02 & 3.49 & 2.52 \\ 
    \bottomrule
    \end{tabular}
    \end{adjustbox}
    \label{table1}
\end{table}
\footnotetext[2]{\url{https://github.com/modelscope/FunASR/tree/main/examples/industrial_data_pretraining/paraformer/}}
\footnotetext[3]{\url{https://github.com/k2-fsa/icefall/tree/master/egs/multi_zh-hans/ASR/}}

\textbf{High-quality close-talk speech segments:} The close-talk speech segments (CTSS) are obtained by segmenting the close-talk recordings using the oracle diarization to minimize speech leakage. By comparing the first and second rows of Table \ref{table1}, we observe that CTSS significantly outperforms GSS signals on both character error rate (CER) and speech quality, demonstrating its reasonability for pseudo labels estimation. 

\textbf{Time alignment:} Due to long-distance signal propagation and the non-synchronization between the close-talk and far-field devices, an time offset exists between the CTSS and the target clean speech (of the reference far-field mixture). To identify this time offset, we apply the classical GCC-PHAT algorithm \cite{dibiase2001robust}, and the time offset ${\tau}'$ is calculated as follows:
\begin{equation}
{\tau}' = \underset{\tau}{\operatorname{argmax}} \frac{1}{2 \pi } \int_{-\infty }^{\infty } \frac{S_{1}(\omega) Y^{*}(\omega)}{\left|S_{1}(\omega) Y^{*}(\omega)\right|} e^{j \omega \tau} d \omega \in \mathbb{R},
\label{eq:t-align}
\end{equation}
where $S_{1}(\omega)$ and $Y(\omega)$ respectively denote the Fourier transforms (FTs) of CTSS signal and the reference far-field mixture. The operator $(\cdot) ^{*}$ represents the complex conjugate, and $\left | \cdot  \right | $ calculates the magnitude. With ${\tau}'$, we shift CTSS signal $s_{1}(t)$ to align with the reference far-field mixture $y(t)$, obtaining $s_{2}(t) $.

\textbf{Level alignment:} The speech signal attenuates rapidly during propagation, resulting in an level-misalignment between $s_{2}(t)$ and the target clean speech, which causes instability in model training. Inspired by the forward convolutive prediction (FCP) \cite{wang2021convolutive}, we perform multi-frame linear filtering (MFLF) on $s_{2}(t)$ to achieve level alignment, where the filter is obtained by solving the following optimization problem:
\begin{equation}
{\boldsymbol{h}}'(f)=\underset{\boldsymbol{h}(f)}{\operatorname{argmin}} \sum_{t} \frac{\left|Y(t, f)-{\boldsymbol{h}}(f)^{\mathrm{H}} \breve{\boldsymbol{S}}_{2}(t, f)\right|^{2}}{\lambda(t, f)} \in \mathbb{C}^{L},
\label{eq:l-align}
\end{equation}
where $Y(t, f)$ represents the short-time Fourier transform (STFT) coefficient of $y(t)$. $\breve{\boldsymbol{S}}_{2}(t, f)=[S_{2}(t, f), \ldots, S_{2}(t-L+1, f)]^{\mathrm{T}} \in \mathbb{C}^{L}$ stacks $L$ T-F units, where $S_{2}(t, f)$ denotes the STFT coefficient of $s_{2}(t)$ and $L$ is set to 2 in our experiment. The operator $(\cdot)^\mathrm{H}$ computes the Hermitian transpose. The term $\lambda(t, f)$, following \cite{wang2021convolutive}, is a weighting factor. Finally, we obtain high-quality pseudo labels that can be directly used for model training: $s_{3}(t) = ISTFT ({\boldsymbol{h}}'(f)^{\mathrm{H}} \breve{\boldsymbol{S}}_{2}(t, f))$, where $ISTFT$ denotes the inverse STFT operation.

\textbf{SNR filter:} In modern supervised learning-based speech enhancement \cite{wang2018supervised,li21g_interspeech}, it is common to impose a lower bound on the SNR of simulated data to prevent extreme samples from adversely affecting model training.  This practice is adopted in this paper, with the difference that since the reference clean speech is unavailable, we estimate the SNR using pseudo labels:
\begin{equation}
\hat{SNR} = 10 \log_{10} \left( \frac{||s_{3}(t)||^{2} }{||s_{3}(t)-y(t)||^{2}} \right).
\label{eq:snr-filter}
\end{equation}
Then, we discard data pairs with $\hat{SNR}$ below -10 dB, and the resulting high-quality data pairs are used for model training.

\subsubsection{Loss function}
\label{section:Loss function}
To improve stability during training on real-recorded data, we propose magnitude constraint adjustable (MCA) loss. The MCA loss is obtained by linearly weighting MSE loss $\mathcal{L} _\mathrm{MSE}$ and cosine similarity loss $\mathcal{L} _\mathrm{COSSIM}$ as follows:
\begin{equation}
\begin{split}
\mathcal{L} _\mathrm{MSE} &= \mathbb{E}_{\boldsymbol{A},\boldsymbol{B}}\left [ || \boldsymbol{A}-\boldsymbol{B} ||_{F}^{2}  \right ],   \\
\mathcal{L} _\mathrm{COSSIM} &= 1-\cos\left ( \boldsymbol{A},\boldsymbol{B} \right ) = 1 - \frac{<\boldsymbol{A},\boldsymbol{B}>_{F} }{|| \boldsymbol{A} ||_{F} || \boldsymbol{B} ||_{F}}, \\
\mathcal{L} _\mathrm{MCA} &= \mathcal{L} _\mathrm{MSE} + \alpha \times \mathcal{L} _\mathrm{COSSIM},
\end{split}
\label{eq:mca}
\end{equation}
where $\boldsymbol{A}$ and $\boldsymbol{B}$ respectively denote the oracle and the predicted magnitude spectrograms. $\cos\left ( \cdot ,\cdot  \right )$ computes the cosine similarity between two magnitude spectrograms, $<\cdot ,\cdot >_{F}$ denotes the Frobenius inner product, and $\alpha$ is a regulator that controls the weight of $\mathcal{L} _\mathrm{COSSIM}$. Unlike $\mathcal{L} _\mathrm{MSE}$, $\mathcal{L} _\mathrm{COSSIM}$ does not strictly force $\boldsymbol{A}$ and $\boldsymbol{B}$ to be identical (i.e., $\mathcal{L} _\mathrm{COSSIM}$ is a softer constraint). Therefore, its introduction makes $\mathcal{L} _\mathrm{MCA}$ more robust.

\section{Back-end}
\subsection{ASR model}
We adopt Paraformer-large \cite{paraformer} as the back-end model, which contains 220M parameters and consists of an encoder, predictor, sampler, and decoder. The continuous integrate-and-fire (CIF) \cite{dong2020cif} based predictor predicts the number of tokens and generates hidden variables. The glancing language model (GLM) \cite{qian2020glancing} based  sampler then generates semantic embeddings to enhance the decoder's ability to model context interdependence. Pre-trained on 60,000 hours of Mandarin speech data, Paraformer demonstrates exceptional performance in Chinese speech recognition, making it well-suited for the MISP-Meeting Challenge.

\begin{figure}
    \centerline{\includegraphics[width=\columnwidth]{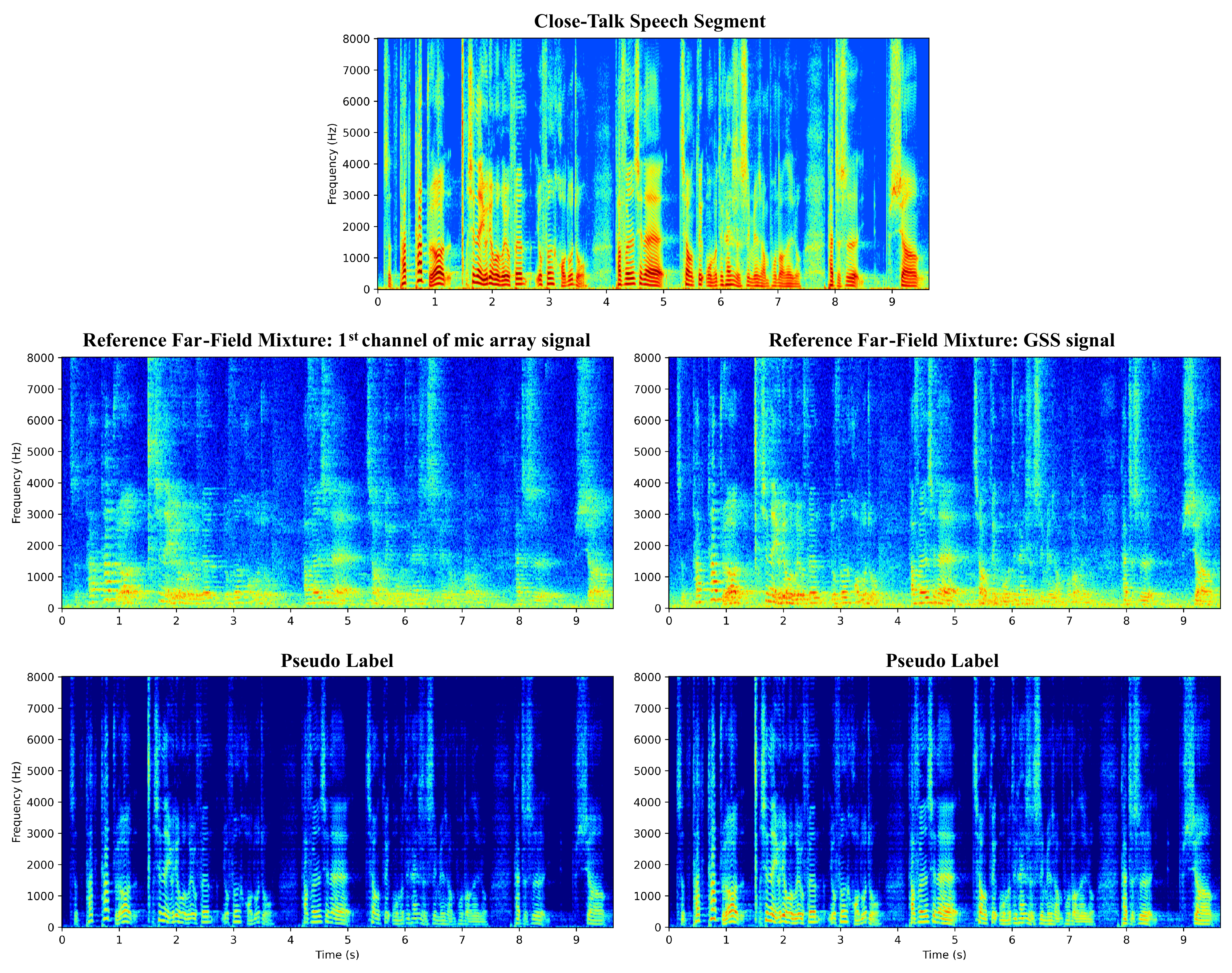}}
    \caption{Spectrogram visualization of close-talk speech segment, reference far-field mixtures, and the corresponding pseudo labels generated by our proposed TLS framework on the MISP-Meeting training set.}
    \label{figure2}
\end{figure}

\begin{figure}
    \centerline{\includegraphics[width=\columnwidth]{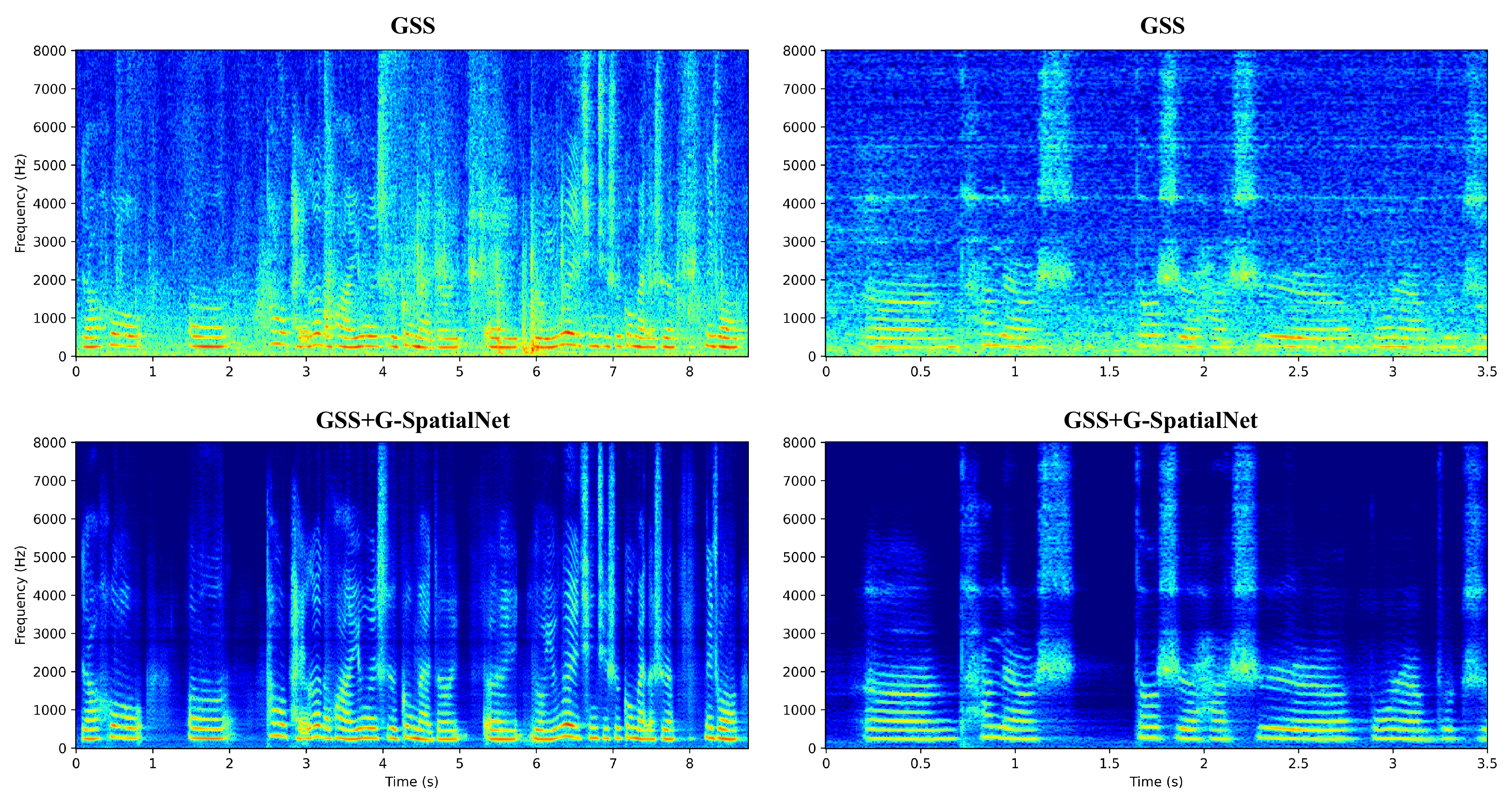}}
    \caption{Spectrogram visualization of far-field mixtures enhanced by GSS and our proposed “GSS+G-SpatialNet” front-end on the MISP-Meeting evaluation set.}
    \label{figure3}
\end{figure}

\subsection{Fine-tuning to adapt to G-SpatialNet}
Since G-SpatialNet aims to minimize the MCA loss rather than the CTC/Attention loss, its enhancement direction may not necessarily lead to a reduction in CER. One effective solution to this issue is to jointly train the G-SpatialNet and Paraformer. However, we are unable to implement this approach due to the limited competition duration. Instead, we adopted a simpler and faster alternative: fine-tuning Paraformer to adapt to G-SpatialNet's output. Fine-tuning consists of two stages: the first stage employs pseudo labels data described in Section \ref{section:Pseudo labels estimation}, and the second stage employs far-field audio data enhanced by “GSS+G-SpatialNet”.

\subsection{Fusion}
We set the maximum number of epochs to 20 during finetuning. To obtain the final model, we averaged the weights of the top five models with the best performance on the development set.

\section{Training data}
\subsection{Speech enhancement}
For G-SpatialNet training, we use only the audio data of MISP-Meeting training set. Specifically, We first apply GSS with the oracle diarization to pre-process the raw 8-channel far-field microphone array signals. Subsequently, pseudo labels are generated using the method described in Section \ref{section:Pseudo labels estimation}. The GSS signals and the corresponding pseudo labels signals are treated as noisy-clean data pairs for training G-SpatialNet.

\subsection{ASR}
For Paraformer fine-tuning, we also use only the audio data of MISP-Meeting training set, as we observe that the introduction of visual modality increases the CER, which will be discussed in Section \ref{section:Data augmentation and visual modality}. The pre-processing of audio data follows the same procedure as described above, and the resulting pseudo labels as well as G-SpatialNet enhanced audio data are used for fine-tuning. Additionally, to explore the impact of data augmentation on pre-trained ASR models, we investigate MISP-Meeting-like Chinese meeting data, namely AliMeeting \cite{yu2022m2met} and AISHELL-4 \cite{aishell4}, and also generate 223 hours of simulated data \footnote{\url{https://github.com/jsalt2020-asrdiar/jsalt2020_simulate/}}.

\section{Result and analysize}
\subsection{Speech enhancement: TLS and G-SpatialNet}
\label{section:TLS and G-SpatialNet}
Fig. \ref{figure2} illustrates an example of pseudo label estimation using our proposed TLS framework. As shown in Fig. \ref{figure2}, the close-talk speech fed to TLS framework time- and level-aligns with the target clean speech of the reference far-field mixtures. Furthermore, we evaluate the CER and speech quality of pseudo labels on the development set. As shown in the third row of Table \ref{table1}, the CER and speech quality of the pseudo label are close to those of the raw close-talk speech. These results indicate that our proposed TLS framework can achieve precise alignment with minimal impact on CER and speech quality, thereby providing high-quality pseudo-labeled data for training modern SE models directly in real-world meeting scenarios.

As illustrated in Fig. \ref{figure3}, G-SpatialNet effectively enhances the GSS signals by removing residual noise and reverberation, thereby improving speech quality. Furthermore, G-SpatialNet significantly reduces CER of ASR back-ends, which is detailed in Section \ref{section:Overall results}.

\begin{table}[]
\caption{CER(\%) $\downarrow$ on the development set with oracle diarization and GSS front-end.}
\begin{adjustbox}{width=\columnwidth}
\label{table:dataaug}
\renewcommand{\arraystretch}{1.2}
\begin{tabular}{ccccc}
\toprule
\textbf{ID} & \textbf{Model} & \textbf{Fine-tuning Data} & \textbf{Duration(h)} & \textbf{CER(\%)} \\ 
\toprule
A1 & Baseline(A) & MISP-Meeting & 119 & 7.78 \\
A2 & Baseline(AV) & MISP-Meeting & 119 & 15.46 \\ 
\hline
B1 & Paraformer & MISP-Meeting & 119 & \textbf{6.31} \\
B2 & Paraformer & simulation+MISP-Meeting & 342 & 6.33 \\
B3 & Paraformer & AliMeeting+MISP-Meeting & 239 & \textbf{6.31} \\
B4 & Paraformer & AISHELL-4+MISP-Meeting & 239 & 6.35 \\
B5 & Paraformer & All & 582 & \textbf{6.31} \\ 
\bottomrule
\end{tabular}
\end{adjustbox}
\end{table}

\subsection{ASR: Data augmentation and visual modality}
\label{section:Data augmentation and visual modality}
In Table \ref{table:dataaug}, the baseline ASR for the audio-only modality (A1) significantly outperform that for the audio-visual modality (A2). The baseline ASR models are fine-tuned from a pre-trained model on the Wenetspeech dataset \cite{wenetspeech}. Fine-tuning with only the audio modality preserves the pre-trained capabilities of the model, while incorporating the visual modality requires re-training the decoder, leading to a loss of some pre-trained abilities and a significant performance drop. To fully leverage the potential of the pre-trained ASR models, we chose to exclude the video modality.

Additionally, as shown in rows 3 to 5 of Table \ref{table:dataaug}, we fine-tune Paraformer on different datasets. Compared to B1, B2-B5 show no performance improvement. We attribute this to the fact that, unlike models trained on small-scale datasets, Paraformer has been pre-trained on a large-scale corpus, which provides strong generalization ability. Therefore, the inclusion of additional, limited datasets not only fails to further enhance Paraformer's generalization but also causes the model to deviate from the MISP-Meeting distribution, leading to performance degradation.

\begin{table}[]
\caption{CER(\%) $\downarrow$ on the development (\texttt{Dev}) and evaluation (\texttt{Eval}) sets with different front-end and back-end configurations. G-SpatialNet\textsuperscript{$\clubsuit$} uses only GSS signals as input, while G-SpatialNet\textsuperscript{$\spadesuit$} utilizes both GSS and 8-channel mic array signals. Paraformer\textsuperscript{$\spadesuit$} is fine-tuned on pseudo labels (Session \ref{section:Pseudo labels estimation}) and then on far-field audio enhanced by G-SpatialNet\textsuperscript{$\spadesuit$}. All other ASR back-ends are finetuned on close-talk and then GSS data.}
\begin{adjustbox}{width=\columnwidth}
\label{table:ovrl}
\renewcommand{\arraystretch}{1.2}
\begin{tabular}{ccccc}
\toprule
\textbf{System} & \textbf{Front-end} & \textbf{Back-end} & \textbf{Dev} & \textbf{Eval} \\ 
\toprule
\multicolumn{3}{c}{Baseline} & 15.46 & 20.1 \\
\hline
S1 & GSS & Paraformer & 6.31 & 10.9 \\
S2 & GSS+G-SpatialNet\textsuperscript{$\clubsuit$} & Paraformer & 6.01 & 10.1 \\
S3 & GSS+G-SpatialNet\textsuperscript{$\spadesuit$} & Paraformer & 5.46 & 9.58 \\
S4 & GSS+G-SpatialNet\textsuperscript{$\spadesuit$} & Paraformer\textsuperscript{$\spadesuit$} & \textbf{5.44} & \textbf{9.52} \\ 
\bottomrule
\end{tabular}
\end{adjustbox}
\end{table}

\subsection{Overall results}
\label{section:Overall results}
Table \ref{table:ovrl} presents the ablation results of our best system. Compared with baseline system, system S1, which employs Paraformer as the ASR back-end, demonstrates a significant performance improvement. Subsequently, by introducing the data-driven G-SpatialNet model, further reduction is achieved even though CER is already relatively low. Specifically, system S3 achieves relative performance improvements of 13.5\% and 12.1\% over system S1 on the development and evaluation sets, respectively, which fully demonstrates the effectiveness of our proposed pseudo labels-based neural speech enhancement method. Additionally, by comparing system S2 and S3, it can be found that the introduction of 8-channel microphone array signal helps to improve the performance of G-SpatialNet, which we attribute to the reuse of spatial information. Finally, by using pseudo labels and G-SpatialNet enhanced audio data to fine-tune Paraformer to align the two, we obtain the optimal system S4. Compared with the baseline, system S4 achieves relative performance improvements of 64.8\% and 52.6\% on the development and evaluation sets, respectively.

\section{Conclusions}
In this paper, we design G-SpatialNet to improve quality of far-field audio. We proposed a novel framework, TLS, which generates high-quality pseudo labels for real-world meeting data, enabling the direct training of SE models in real-world scenarios. Furthermore, we explore how fine-tuning strategies, data augmentation, and multimodal information can further boost the performance of pre-trained ASR models. Finally, our system achieved CERs of 5.44\% and 9.52\% on the development and evaluation sets of the MISP-Meeting Challenge, securing second place.

\section{Acknowledgments}
This work was supported in part by the National Natural Science Foundation of China under Grants 62371407 and 62276220, and the Innovation of Policing Science and Technology, Fujian province (Grant number: 2024Y0068).

\bibliographystyle{IEEEtran}
\bibliography{mybib}

\end{document}